\documentclass[pra,floatfix,twocolumn,superscriptaddress]{revtex4}
\usepackage{times,graphicx,bbm,amsmath,amssymb}
\usepackage{epsfig,color}
\usepackage[colorlinks=true,citecolor=blue]{hyperref}

\def \AD{\textcolor{black}}

\usepackage{pifont}

\newcommand{\ket}[1]{\vert#1\rangle}

\begin{document}

\title{Quantum enhanced estimation of optical detector efficiencies}

\author{Marco Barbieri} 
\affiliation{Dipartimento di Scienze, Universit\`a degli Studi Roma Tre, Via della Vasca Navale 84, 00146, Rome, Italy}
\affiliation{Clarendon Laboratory, University of Oxford, Parks Road, OX1 3PU, Oxford, UK}
\author{Animesh Datta}
\affiliation{Clarendon Laboratory, University of Oxford, Parks Road, OX1 3PU, Oxford, UK}
\author{Tim J. Bartley} 
\affiliation{Clarendon Laboratory, University of Oxford, Parks Road, OX1 3PU, Oxford, UK}
\affiliation{National Institute of Standards and Technology, 325 Broadway, Boulder, CO 80303, USA}
\author{Xian-Min Jin}
\affiliation{Clarendon Laboratory, University of Oxford, Parks Road, OX1 3PU, Oxford, UK}
\affiliation{Department of Physics and Astronomy, Shanghai Jiao Tong University, Shanghai 200240, China}
\author{W. Steven Kolthammer}
\author{Ian A. Walmsley}
\affiliation{Clarendon Laboratory, University of Oxford, Parks Road, OX1 3PU, Oxford, UK}

\begin{abstract}

Quantum mechanics establishes the ultimate limit to the scaling of the precision on any parameter, by identifying optimal probe states and measurements. While this paradigm is, at least in principle, adequate for the metrology of quantum channels involving the estimation of phase and loss parameters, we show that estimating the loss parameters associated with a quantum channel and  a realistic quantum detector are fundamentally different. While Fock states are provably optimal for the former, we identify a crossover in the nature of the optimal probe state for estimating detector imperfections as a function of the loss parameter. We provide explicit results for on-off and homodyne detectors, the most widely used detectors in quantum photonics technologies.

\end{abstract}

\maketitle

\section{Introduction}
Achieving measurements that approach the ultimate limits to precision is a fundamental, and stimulating, challenge for experimental sciences, and efforts might be rewarded with glimpses of novel effects or solid confirmation of existing paradigms at unprecedented scales. Quantum mechanics dictates the essential limits to measurement, and it is often the case that ultimate precision is achieved by preparing a measurement probe in a state exhibiting genuine quantum features, such as entanglement~\cite{Holland93,Giovannetti04,Lee02,Higgins07,Dorner09} \AD{, and squeezing~\cite{GEO600,LIGO}. Optical metrology has shown particular promise in this regard; obtaining a quantum advantage in the precision of phase estimation is within reach with some improvements in present technology~\cite{Datta11,Yone,Matthews13}.} Concepts from quantum parameter estimation readily apply to quantum processes. A familiar example is given by the estimation of a lossy bosonic channel, which could describe, for example, an optical transmission line~\cite{Monras07,Adesso09,Inver11}. The relevant parameter, the \AD{transmissivity} of the channel, can be estimated \AD{with quantum limited} uncertainty by using Gaussian probe states, such as coherent states, and photon counting~\cite{Monras07, Pinel13}.

A key aspect of quantum estimation is the characterization of a measurement apparatus itself. Increasingly efficient particle detectors are being leveraged in a wide range of fields, from observational astronomy to quantum information science, and the task of precisely determining the efficiency of these detectors is necessary for their optimal use~\cite{Fermi2009, Marsili13}. Inefficient detectors can be modelled as a channel loss followed by a perfect detector; such a description is commonplace in the analysis of quantum optical experiments~\cite{Bachor04}, and applies to detectors of electromagnetic radiation in general. 

When estimating the inefficiency of a detector, the POVM is fixed to that implemented by the ideal detector. This is fundamentally different from the estimation of any other parameter, where one is free to choose the measurement.  This restriction in the choice of the measurement may in general prevent the attainment of the quantum Cram\'er-Rao bound~\cite{Braunstein94, Paris09}. This makes the estimation of the efficiency of a detector conceptually different from other quantum-limited estimation problems.

In this paper, we analyse the use of Fock states as resource for the estimation of detector efficiency applied to the two most common optical detection apparatus: on-off counting and homodyne detection. While both play a vital role in quantum optics experiments, the former has also appeal in astronomical observations. Our main finding is the existence of regions in which Fock states provide superior performances to coherent states normally used for calibration. Crucially, this is in the region of very high efficiencies where the forefront of detector development lies. We also stress the advantages that Fock states are able to provide with respect to the case of a lossy channel estimation. Our analysis of the performance of quantum light in estimating detector efficiencies with a higher precision concerns those researchers looking for applications of these detectors in their studies. These include, for instance, the use of single-photon correlation measurement of celestial bodies in astrometry, and the employ of single-photon detection for precise imaging in biology.

\section{On-off detectors} 

An extensively adopted detection scheme uses on/off (Geiger-mode) detectors, which are able to tell the presence of photons in the measured field, but can not determine the number of photons. Limitations to the detection efficiency arise from the physical process of light absorption and conversion of the energy into an electrical signal, as well as any loss on the detection setup. Here we address the task of estimating detection efficiency, and show an advantage of quantum optical probe states over classical probes with the same energy. As we show, this advantage is rapidly lost in the presence of dark counts.


We perform our analysis in terms of the Fisher information~\cite{Braunstein94,Paris09} $F(\eta){=}\sum_x \left(\partial_\eta p(x;\eta)\right)^2/p(x;\eta)$  for the efficiency $\eta$, in which the sum is over each possible detector outcome $x$ occuring with probability $p(x;\eta)$. The Fisher information bounds the uncertainty, quantified by the variance $\Delta^2\eta$, attained after $M$ repetitions of the experiment through the Cram\'er-Rao bound $\Delta^2\eta\geq1/(M F(\eta))$ ~\cite{Paris09}. Since the detection scheme is fixed by definition in our problem, there is no possibility of invoking the quantum Cram\'er-Rao bound, i.e. the Cram\'er-Rao optimised over all possible measurements. Prior works have used Bayesians estimators for example to saturate this bound in actual experiments, using a modest number of samples~\cite{Genoni}.  

We begin with the noiseless case in which there are no dark count events. For an $n$\AD{-} photon Fock state, the probabilities of a click (on) or a no-click (off) event give the following expression for the Fisher information
\begin{equation}
F_n(\eta)=\frac{n^2(1-\eta)^{n-2}}{1-(1-\eta)^n}.
\end{equation} 

\begin{figure}[t!]
\includegraphics[viewport = 0 0 600 450, clip, width=\columnwidth]{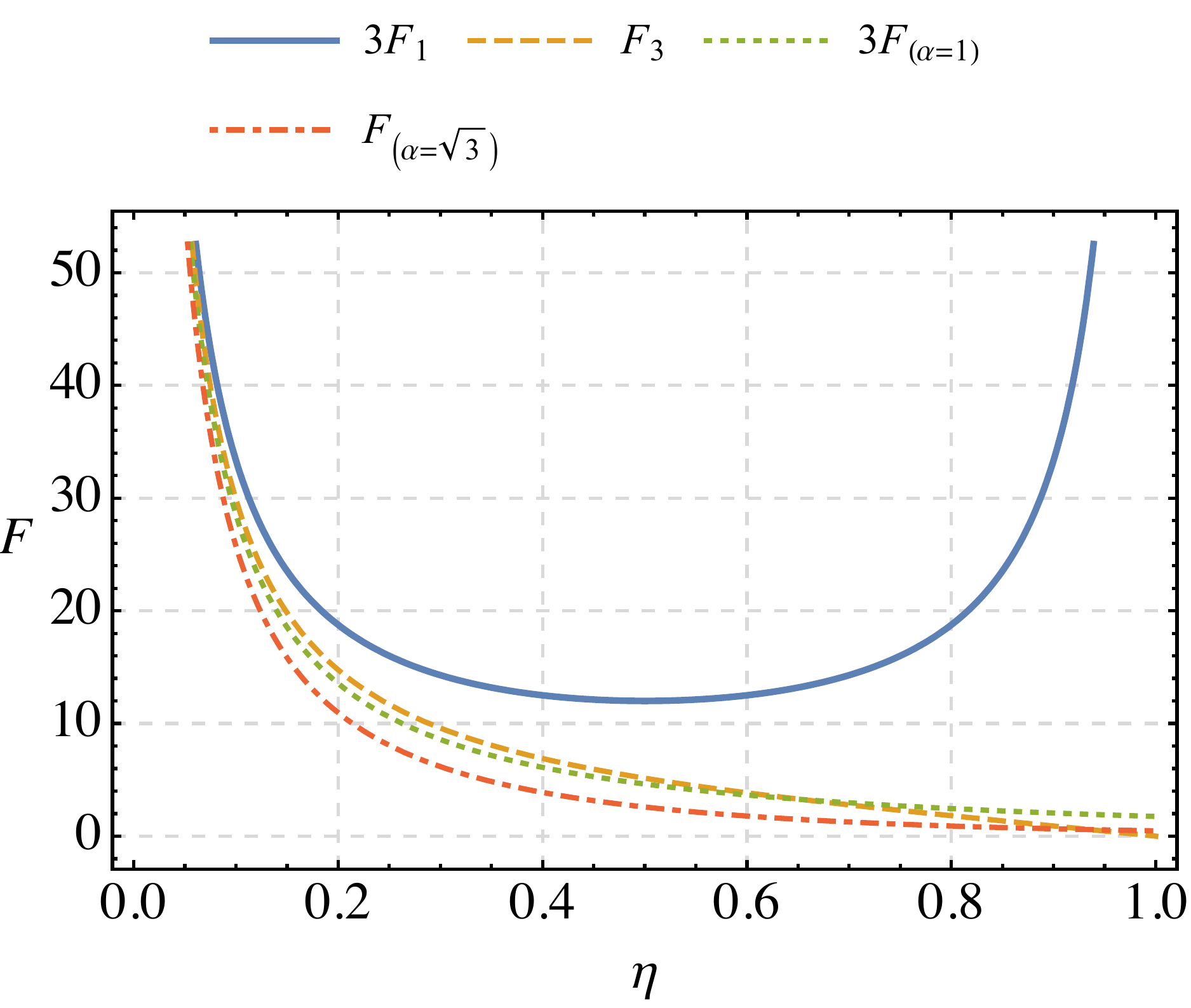}
\caption{Comparison of the Fisher information for the case of the on/off detector with either Fock-state probes ($n{=}1$, solid blue line; $n{=}3$  dashed gold line) or coherent states ($|\alpha|^2{=}1$,  dotted green line; $|\alpha|^2{=}1$  dot-dashed red line). The comparison is carried for the same energy so to illustrate the statistical reduction.}
\label{fig:Fock}
\end{figure}

It is instructive to analyse the limiting cases of low and high efficiencies. In the former case, the leading term scales as $F_n(\eta){\sim}n/\eta$. The achievable precision thus increases with $n$; however the scaling is such that the same advantage is obtained by running the experiment $n$ times with single photons. This is illustrated in Fig. \ref{fig:Fock} in which we plot the Fisher information for $n{=}1$ and $n{=}3$ at fixed probe energy, i.e., we compare $F_3(\eta)$ and $3F_1(\eta)$. In the low efficiency region, the two curves match. Outside of this region, a wiser use of the energy is preparing $n$ times as many single photons, rather than concentrating it on a single $n$ photon state; in the appendix, we demonstrate the optimality of this strategy.

In the opposite limit of a highly efficient detector, $\eta{\simeq}1$, the leading term is $F_n(\eta)\sim n^2(1-\eta)^{n-2}.$ The quadratic increase is compensated by the exponential term, except in the cases $n{=}1$, and $n{=}2$. For single photons, a divergence is obtained for $\eta{\simeq}1$, which makes them more suitable probe states. We notice that this is in contrast with known results for loss estimation, for which a scaling $\Delta^2\eta{\sim}n^{-1}$ can be achieved with $n$-photon states~\cite{Adesso09}. The fact that the detector fixes the measurement prevents one from choosing the  POVM for the estimation of the loss parameter as the optimal given by the symmetric logarithmic derivative. 


The advantages of quantum light can be assessed by inspecting the performance of coherent states; a suitable comparison can be made by considering probe states with the same average energy.The Fisher information for a coherent state $\ket{\alpha}$ is 
\begin{equation}
F_{(\alpha)}(\eta)=\frac{|\alpha|^4}{1-e^{\eta|\alpha|^2}}
\end{equation} 
The comparison is carried out in Fig.\ref{fig:Fock} for both three uses of a coherent state with $|\alpha|^2{=}1$ and a single use of $|\alpha|^2{=}3$. In the low efficiency limit, the scaling with the energy is the same as  with Fock states \AD{$F_{(\alpha)}(\eta){\sim}|\alpha|^2/\eta$}, making the performance of a coherent state closely match the one of a single photon. \AD{Elsewhere, particularly in the high efficiency regime, coherent states are always outperformed by single photons.} Although to a lesser extent than with Fock states, the advantage of multiple, less energetic probes appears with coherent states as well.  
 

\begin{figure}[b!]
\includegraphics[viewport = 0 0 550 400, clip, width=\columnwidth]{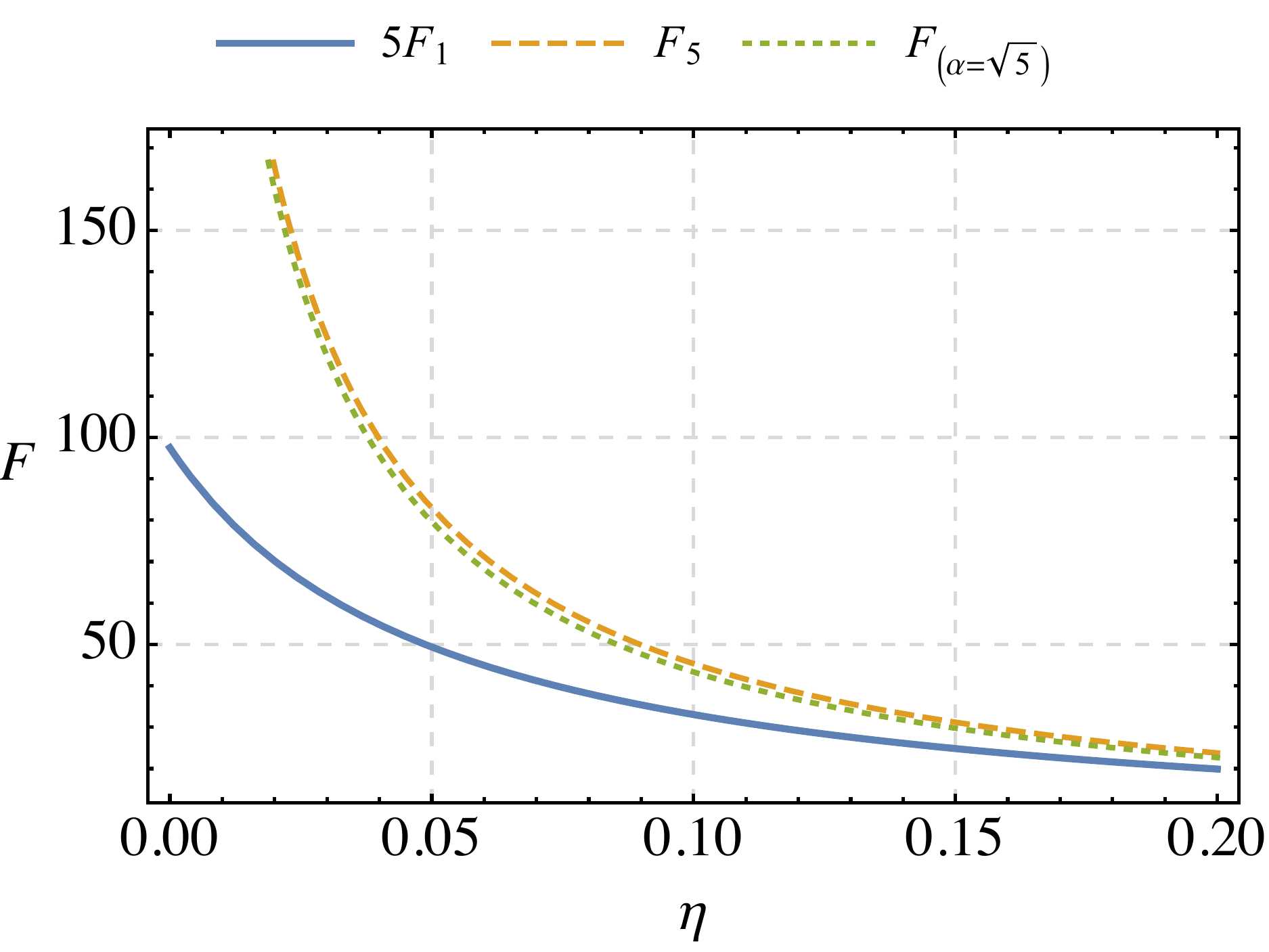}
\caption{Comparison of the Fisher information for a noisy on/off detector with  $\delta{=}0.05$: $n{=}1$ Fock state (solid blue line), $n{=}5$ Fock state (dashed gold line), and a coherent state with $|\alpha|^2{=}5$ (dotted green line). As in Fig.1, the analysis considers fixed energy.}
\label{fig:noise}
\end{figure}

Most practical detectors are affected by dark counts, and we take that into account in the following. A dark count corresponds to an event in which the detector response indicates the presence of photons, despite an input state of vacuum. The probability of a dark count can typically be measured with very low uncertainty, so we assume this error to be negligible and proceed with a single-parameter estimation of $\eta$. This process is modelled as a thermal state reducing the probability of no counts in presence of vacuum by $e^\delta$~\cite{Audenaert09}, which modifies the expression for the Fisher information as 
\begin{equation}
\begin{aligned}
F_n(\eta,\delta)=\frac{n^2(1-\eta)^{n-2}}{e^\delta-(1-\eta)^n}\AD{,}\\
F_{(\alpha)}(\eta,\delta)=\frac{|\alpha|^4}{1-e^{\delta+\eta|\alpha|^2}}
\end{aligned}
\end{equation}
Optimality of different states at low efficiency $\eta{\sim}0$ can be assessed by considering a truncated expansion:
\begin{equation}
\begin{aligned}
F_n(\eta,\delta)\sim\frac{n}{\eta+\frac{e^\delta-1}{n}}\\
F_{(\alpha)}(\eta,\delta)\sim\frac{|\alpha|^2}{e^{\delta}\eta+\frac{e^\delta-1}{|\alpha|^2}}
\end{aligned}
\end{equation}
Interesting features become evident in this limit. Fock states with $n$ photons, and $n$ repetitions of single photons are not equivalent: the $n$-photon state suppresses the additional cost term $e^\delta{-}1$ by a factor $n$, hence large Fock state perform fundamentally better. However, coherent states present a very similar behaviour, with a slight difference coming from the fact that $\eta$ is modified by a factor $e^{-\delta}$, which, in many relevant cases can be close to unity. A numerical comparison of these approximate expressions is shown in Fig.~\ref{fig:noise}, for the case $\delta{=}0.05$: the advantage of the $\ket{n{=}5}$ state over five replicas of a single photon is evident, while the relative advantage with respect a coherent state is much less pronounced.    

The methods employed above can also be applied to the case of $K$-outcome detectors,  which can resolve $0,1,..K-1$ photons, but can not distinguish  $K$ from $K{+}1$ photons. This is a limitation typically encountered in photon-number resolving detectors, such as transition-edge sensors. We also consider the simple case in which the efficiency is independent on the incoming photon number, i.e. the probability of an outcome $k<K$ given the input $\ket{n}$ is simply $p_k={n \choose k}\eta^k(1-\eta)^{n-k}$, where $n\geq K$. From the expression of the Fisher information reported in the appendix, we found that the Fock state with $n=K$ delivers the largest information for all values of $\eta$: $F_{K;K}(\eta)=\frac{K-1}{\eta(1-\eta)}$. The behaviour is then analogous to that of the on/off detector. A more refined treatment requires to consider a general expression in which the probabilities $p_k$ are generated by a stochastic matrix taking into account the fact that the efficiency might depend on the particular outcome.

\section{Homodyne detectors.} 

We now address the case of the homodyne detector, an example of Gaussian measurement. Here we consider the Fisher information associated with the probability of detecting a quadrature amplitude $q$. As above, we compare the use of Fock states and coherent states. In the latter case, we assume, for simplicity, that the coherent state is in phase with the local oscillator of the homodyne detector. When accounting for the detection efficiencies the distributions become:
\begin{equation}
\begin{aligned}
&p_n(q;\eta){=}\frac{e^{-2q^2}}{\pi}\sum_{m{=}0}^n \binom{n}{m} {\eta^m(1-\eta)^{n{-}m}}\frac{h_m^2(q)}{2^m m!}\frac{h_{n{-}m}^2(q)}{2^{n{-}m}(n{-}m)!}\\
&p_{(\alpha)}(q;\eta)=\frac{e^{-(q-\sqrt{2\eta}\alpha)^2}}{\sqrt{\pi}}
\end{aligned}
\end{equation}
where $h_m(q)$ is the $m$-th Hermite polynomial, and we set the standard quantum noise to 1/2.

\begin{figure}[th]
\includegraphics[viewport = 60 0 600 410, clip, width=\columnwidth]{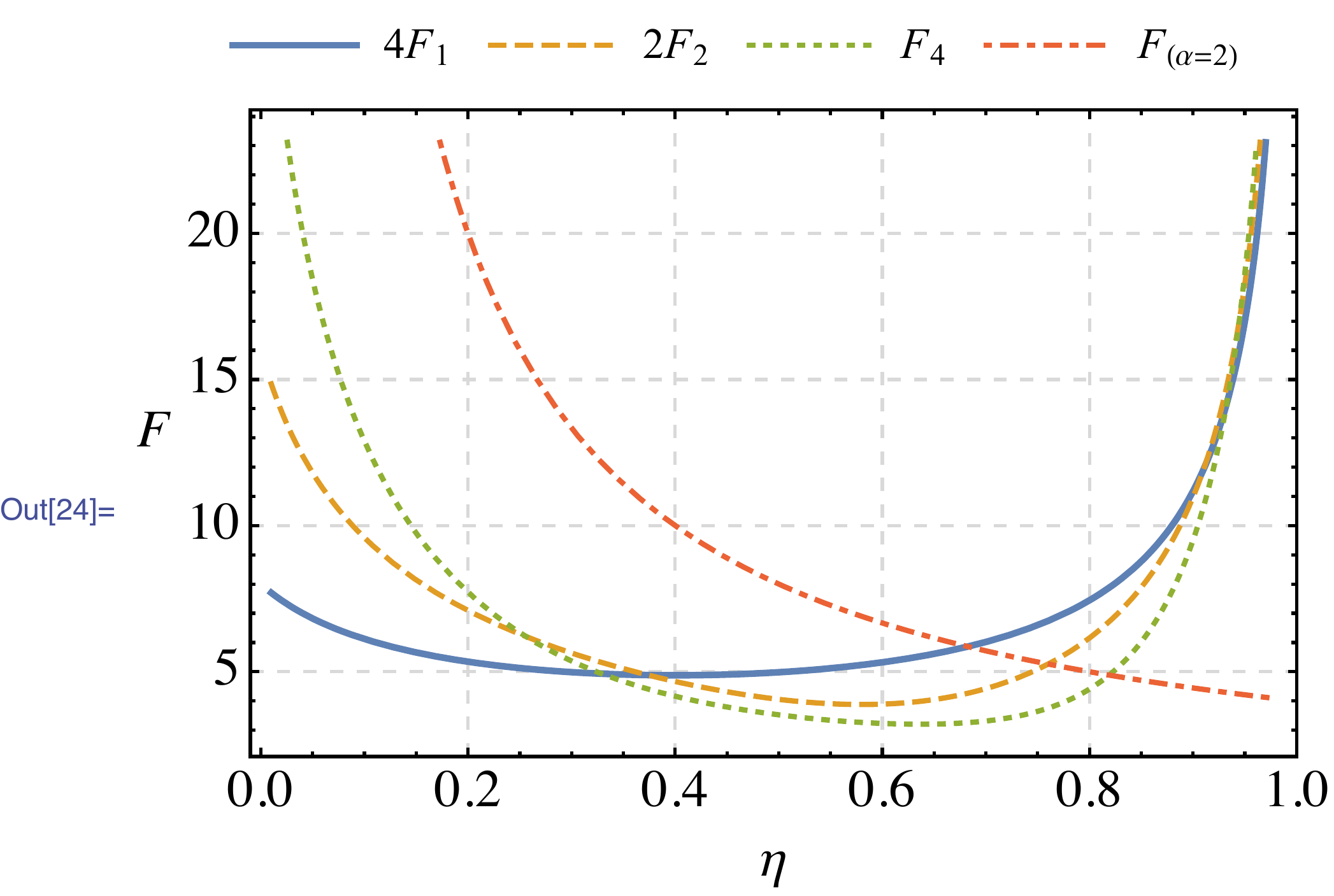}
\caption{Comparison of the Fisher information for the homodyne detector: $n{=}1$ Fock state (solid blue line), $n{=}2$ Fock state (dashed gold line),  $n{=}4$ Fock state (dotted green line), and a coherent state with $|\alpha|^2{=}4$ (dot-dashed red line).}
\label{fig:HD}
\end{figure}

For the coherent state we obtain an expression for the Fisher information: $F_{(\alpha)}(\eta){=}\alpha^2/\eta$~\cite{Pinel13,Monras}. For the case of Fock states it has not been possible to obtain closed forms for the integrals on $x$ leading to the Fisher information; therefore, we have performed these integrals numerically for a set of chosen values of $\eta$, and then interpolated. The comparison is reported in Fig.~\ref{fig:HD}, where we show the behaviour at a fixed energy with $4$ photons distributed in either $n{=}1$, $n{=}2$ or $n{=}4$ Fock states, in addition with a coherent state with $|\alpha|^2{=}4$. We then compare two different strategies. When using coherent states, we aim at distinguishing a displacement of the Gaussian quadrature distribution, with respect to the vacuum; on the other hand, we rely on non-Gaussian features of the distributions when using Fock states~\cite{Raymer09}.

For high efficiencies $\eta \sim 1$, Fock states perform better than coherent states. Intuitively, in this limit, the rapidly-varying features originating from the Fock states enable a measurement that is more sensitive to inefficiency than the smooth Gaussian distribution resulting from coherent states. In the limit of very high efficiency, the more baroque the quadrature distribution, the more precise the measurement. Such features, though, are rapidly smoothed out as the efficiency decreases. \AD{For values of efficiencies ranging from $\eta{\simeq}0.6$ to 0.95, the single photon state might still offer some advantage over a coherent state, while higher photon numbers are less appealing as resources.}

In the opposite limit of low efficiencies, higher-number states become more \AD{favourable} than a single-photon state, due to the presence of more non-Gaussian features in the distributions; however, the best option appears to be the estimation of the displacement due to the coherent state. Our analysis also applies in the presence of electronic noise, which can be encompassed in an effective efficiency~\cite{Appel07}.

\section{Practical considerations.}

Single photons are often produced by spontaneous parametric down conversion~\cite{Elsaman11}. The photon pair production mechanism is employed in such a way that one particle acts as a herald for a second photon emitted on a correlated mode. Although the generation occurs in pairs to satisfy energy conservation, imperfect mode matching makes it possible to herald photons with a limited efficiency $\xi$, which then represents the average photon number on the mode. Such limited heralding efficiency can be responsible for a loss of advantage in the calibration of detector efficiency; while the use of such a scheme has been known to be useful for for absolute calibration~\cite{Klyshko80,Migdall96,Brida02}, here we analyse its performance in terms of use of resources.  

\begin{figure}[th]
\includegraphics[viewport = 0 0 600 410, clip, width=\columnwidth]{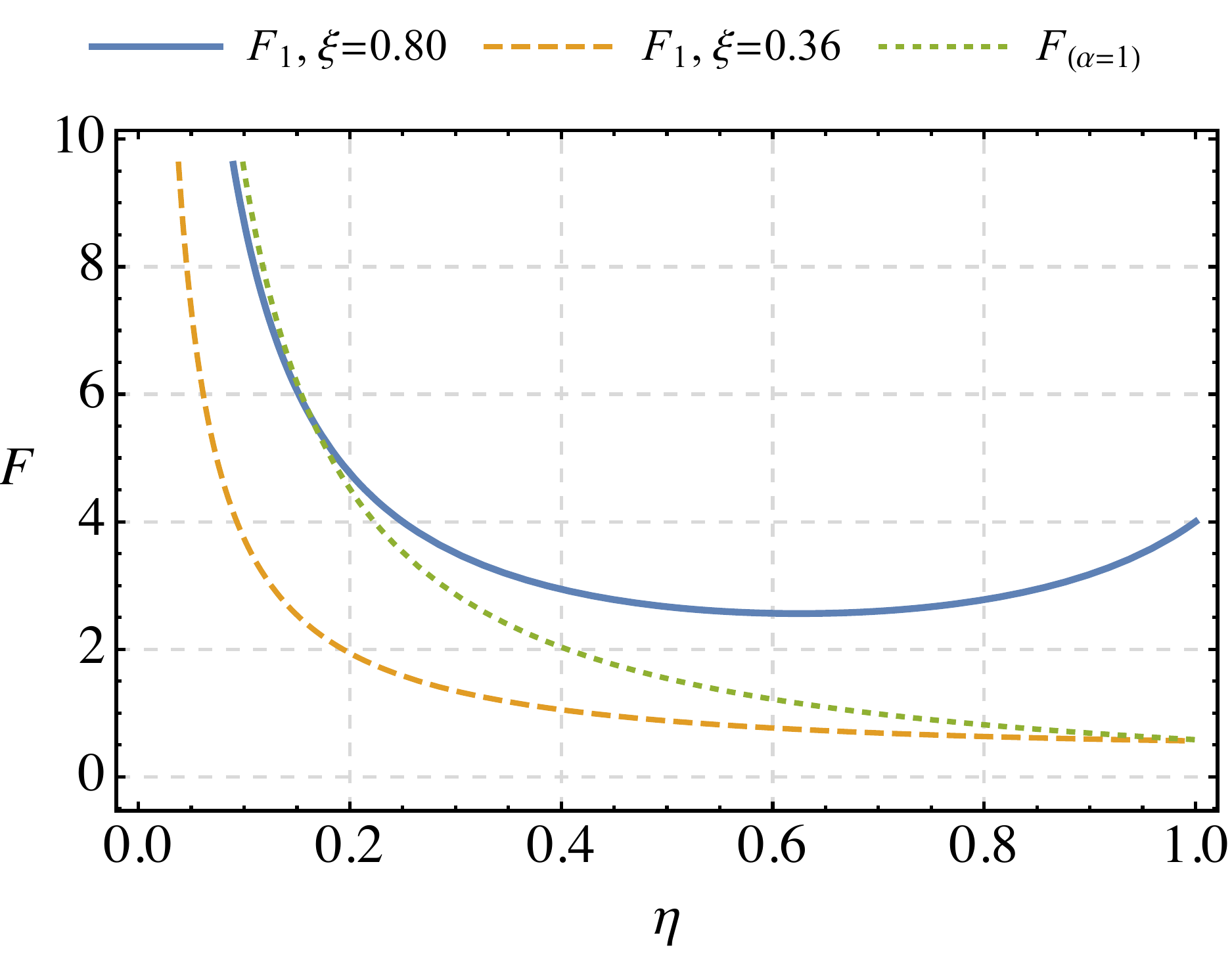}
\caption{Comparison of the performance of a single photon in measuring the efficiency of the on-off detector in presence of a limited heralding efficiency $\xi$: $\xi{=}0.80$ (solid blue line), $\xi{=}1/e\simeq 0.36$ (dashed gold line). The performance of a coherent state $|\alpha|^2{=}1$ is also reported (dotted green line).}
\label{fig:heralding}
\end{figure}

Our main findings are illustrated in Fig.~\ref{fig:heralding} for the case of an on-off detector. One can tolerate limited heralding efficiency while still preserving an advantage with respect to the use of coherent states with an average photon number of one -- the comparison is thus not carried out at fixed energy, to understand whether it is profitable to generate quantum resources. Remarkably, for high efficiencies the fact that the photon statistics is not Poissonian allows to tolerate heralding efficiencies as low as 0.36 to retain an advantage over coherent states; values of approximately 0.8 have recently been reported~\cite{Spring13,Harder13}. The qualitative behaviour for the case of homodyne detection has been found to be similar, albeit the requirements on the heralding efficiencies are more demanding: the minimal value for appreciating an advantage over a coherent state with $|\alpha|{=}1$ has been found to be $\xi\simeq0.765$.

As to the practical implementation, we observe that, for an accurate estimation, the value of $\xi$ needs to be calibrated beforehand, while the efficiency of the detector does not enter the calculations, since it only sets the generation rate. Also, the presence of high-order terms in the emission might affect the precision, however, one can limit their presence by a continuous-wave pump~\cite{Brida12}.  

\section{Conclusions.} 

We have presented a detailed analysis of the estimation of the efficiency of photonic detectors using Fock states, and the benchmarking of their performance against coherent states of the same average photon number. We have found that single photons are close to the optimal performance for both an on-off detector and a homodyne detector for high efficiencies; in the regime of low efficiencies, however, their advantage is modest, if not inferior to that of coherent states. The choice of such probe states is dictated by the feasibility of their sources. Future directions will also include a full characterisation of photon-resolving detectors, which require a multi-parameter approach, as the whole matrix linking the 
input photon number and the output click number must be taken into account.

\section*{Acknowldgements.} We thank Joshua Nunn, Peter Humphreys, Gaia Donati, and Mihai Vidrighin for discussion and comments. M.B. is supported by a Rita Levi-Montalcini contract of MIUR.  AD is supported an EPSRC Fellowship (EP/K04057X/1). X.M.J. is supported by an EU Marie-Curie Fellowship (PIIF-GA-2011-300820). WSK is supported by an EU Marie-Curie Fellowship (PIIF-GA-2011-331859). This work was supported by the EPSRC (EP/H03031X/1, EP/ K034480/1), the European Commission project SIQS, the Air Force Office of Scientific Research (European Office of Aerospace Research and Development). 

\vskip 5mm

\section*{Appendix}
\emph{Proof that $\ket{1}$ is optimal for on-off detectors.} For any Fock state $\ket{n}$, we found that the Fisher information is bound as $F_n(\eta)\leq n F_1(\eta)$, which descends from Eq.(1). We then consider the use of  $M$ photons overall for estimating the efficiency $\eta$, either split into $M$ single photons, or in the state  $\ket{\psi}{=}\sum_{j\geq0}x_j\ket{j}$ with average photon number $\bar n{=}\sum_{j\geq 0}|x_j|^2 j$, with $~M/{\bar n}$ repetitions. Noticing that the POVM of the detector is insensitive to the coherences in the Fock basis, we have: $F_{\psi}(\eta)\leq \sum_{j\geq 0}|x_j|^2 j F_j(\eta)\leq {\bar n} F_1(\eta)$, due to the convexity of the Fisher information. If we now also consider the different number of copies available, one can infer the optimality of single photons. 

\emph{General expression of the Fisher information for a $K$-outcome detector.}  Here we assume that the detector can distinguish up to $K-2$ photons, with an additional outcome that flags events with a higher photon number without resolving them. When considering a Fock state $\ket{n}$, with $n\geq K$ one finds:
\begin{equation*}
\begin{aligned}
F_{n;K}(\eta)=\sum_{k=0}^{K-2} {n \choose k} (1-\eta)^{n-k-2}\eta^{k-2} (k-n\eta)^2 + \frac{(\partial_\eta p_K)^2}{p_K},
\end{aligned}
\end{equation*}
with $p_K=1-\sum_{k=0}^{K-2}  {n \choose k} (1-\eta)^{n-k}\eta^{k}.$

\end{document}